\documentclass[]{article} 
\usepackage{emulateapj}
\usepackage{graphicx}
\usepackage{multirow}

\advance \voffset by -0.5cm\relax

\usepackage{color}

\def\msun{{\rm ~M}_{\odot}}
\def\rsun{{\rm ~R}_{\odot}}

\def\mpy{{\rm ~M}_{\odot} {\rm ~yr}^{-1}}
\def\zsun{{\rm ~Z}_{\odot}}
\def\mps{{\rm ~M}_{\odot} {\rm ~s}^{-1}}

\begin{document}

\title{High Mass X-ray Binaries: Future Evolution and Fate}

 \author{Krzysztof Belczynski\altaffilmark{1,2},
         Tomasz Bulik\altaffilmark{1}, Chris L. Fryer\altaffilmark{3,4,5}}

\affil{
     $^{1}$ Astronomical Observatory, University of Warsaw, Al.
            Ujazdowskie 4, 00-478 Warsaw, Poland\\
     $^{2}$ Center for Gravitational Wave Astronomy, University of Texas at
            Brownsville, Brownsville, TX 78520, USA\\
     $^{3}$ Computational Computer Science Division, Los Alamos National
            Laboratory, Los Alamos, NM, USA\\
     $^{4}$ Department of Physics, The University of 
            Arizona, Tucson, AZ, USA\\
     $^{5}$ Department of Physics and Astronomy, The University of New 
            Mexico, Albuquerque, NM, USA\\
 }
 
\begin{abstract}

BH-NS and BH-BH systems are among the most promising gravitational
wave sources detectable by advanced LIGO/VIRGO and the Einstein
Telescope.  Although the rates of these systems may be above those of
NS-NS mergers, BH-NS and BH-BH systems are difficult to detect, and
thusfar none have been observed.  But the progenitors of BH-NS and
BH-BH binary systems may have been observed, in the form of High-Mass
X-ray Binaries (HMXBs).  In this paper, we continue work studying
these potential progenitors of these important gravitational wave
sources.  In the first two papers of the series, we have demonstrated
that IC10 X-1 and NGC300 X-1 are direct progenitors of BH-BH systems
and that Cyg X-1 may form, alas with a very low probability, a BH-NS
system. Here, we analyze the Galactic binaries GX 301-2, Vela X-1,
XTEJ1855-026, 4U1907+09, Cir X-1 and extra-galactic LMC X-1, LMC X-3,
M33 X-1. In each case, we find that the future evolution will not allow
the formation of a BH-NS nor a BH-BH system. Most of these binaries
will soon merge in the common envelope phase, with a compact object
sinking into a helium-rich core of a stellar companion.  This
``helium-merger'' may be a progenitor for long duration gamma-ray
bursts (GRBs).  Based on the observed HMXB population, the rate of
helium-mergers may make up a sizable fraction of long-duration GRBs.
Due to this high number of potential GRB progenitors, a chance that a
Galactic HMXB has caused one of the recent major mass extinction
events is significant ($\sim 10-20\%$).

\end{abstract}


\section{Introduction}

Binary compact mergers are the most promising sources for ground-based
gravitational wave detectors such as LIGO or VIRGO.  These
observatories are currently being upgraded and are scheduled to begin
operating in the 2015 time frame.  With the heightened sensitivity of
these upgraded detectors, the questions surrounding gravitational wave
detection are evolving from ``will we detect gravitational waves?''
to ``when will we detect gravitational waves?'' and ``what type of
merging remnant will dominate the detections?''.  So far, astronomers
have only observed double neutron stars (NS-NS), detecting the pulsar
in radio frequencies of electromagnetic spectrum (e.g., Lorimer 2008).
Kim, Kalogera \& Lorimer (2010) assessed empirical NS-NS inpiral rates
and it appears that the detection of these events in gravitational
radiation is unavoidable once the advanced instruments reach their
design sensitivity.  But the upgraded LIGO and VIRGO systems should
also be sensitive to other types of double compact objects, namely
black hole neutron star systems (BH-NS) and double black holes
(BH-BH).  Unlike NS-NS binaries, these binary systems have yet to be
observed.  In principle, these types of double compact objects are
within reach of radio and microlensing surveys. Dong et al. (2007)
reported the first potential detection of BH-BH system, but allowed
for an alternative interpretation of the {\tt OGLE-2005-SMC-001}
lensing event.  Population synthesis results based on theoretical
evolutionary considerations indicate that BH-NS and BH-BH systems are
expected to form and populate the local universe (e.g., Lipunov, Postnov
\& Prokhorov 1997; Bethe \& Brown 1998; De Donder \& Vanbeveren 1998;
Bloom, Sigurdsson \& Pols 1999; Fryer, Woosley \& Hartmann 1999;
Nelemans, Yungelson \& Portegies Zwart 2001, Belczynski, Kalogera \&
Bulik 2002, Voss \& Tauris 2003). In particular, it was recently
argued that BH-BH systems will dominate the gravitational radiation
inspirals (Belczynski et al. 2010b).

At {\em Warsaw Observatory} we have undertaken a program to provide
empirical constraints on the existence of these undetected classes of
compact binary systems.  Bulik, Belczynski \& Prestwich (2011) studied
the future evolution of two extra-galactic X-ray binaries hosting
massive black holes with Wolf-Rayet companions: IC10 X-1 and NGC300
X-1. It was argued that these binaries will produce massive BH-BH
systems. Based on a simple rate estimate, it was shown that the detection
of such massive BH-BH inspirals by gravitational detectors is likely
even in the existing data of the initial LIGO and VIRGO.  Belczynski,
Bulik \& Bailyn (2011) have studied the famous Galactic binary Cyg X-1
hosting a massive black hole and an O type companion. It was found
that the most likely fate of this binary is disruption during the
supernova explosion that forms a neutron star out of the O type
companion. Cyg X-1 has only a small chance of forming a close BH-NS
system. The rate of BH-NS systems inferred from Cyg X-1 is too low
to make them likely sources for advanced gravitational radiation
detectors.  However, it was noted that this finding represents only a
lower limit on the detection of BH-NS systems as these objects may
form along channels that do not involve Cyg X-1 like stage.

In this study we discuss the remaining Galactic and extra-galactic high mass 
X-ray binaries (HMXBs) that are putative progenitors for the BH-NS or BH-BH 
systems. The binaries under consideration host already one compact object, a
neutron star or a black hole, and a massive companion that can in principle form 
the second compact object. We chose only those binaries that have rather well 
established system parameters. We do not have any strict definition for our 
sample as we have chosen (rather subjectively) binaries that either appear to 
have a chance to form BH-NS or BH-BH system, or were brought to our attention by 
someone from the compact object community. In some cases, we had to perform 
detailed calculations to establish the fate of a given system. However, in other 
cases it was enough to search literature for a better parameter estimate or just 
to reiterate the previously presented arguments to confirm already proposed  
conclusions.  As we will show, it appears that besides considered earlier binaries 
IC10 X-1, NGC300 X-1 and Cyg X-1, none other is likely to form BH-NS nor BH-BH 
system. 

In \S\,2, we review the fates of each system individually.  In many cases, the 
binary system is expected to merge.  This merger of a compact remnant with a 
massive companion has been proposed as a progenitor of long-duration GRBs 
(Fryer \& Woosley 1998).   We discuss the implications of these systems 
on the GRB population in \S\,3.

\section{Case Study}

\subsection{GX 301-2/BP Cru: CE merger}

This system consists of a neutron star (NS) and a very massive
companion. The orbital period was measured at $P_{\rm orb}=41.5$d, the
orbit is eccentric $e=0.46$, the NS was estimated to have high mass
$M_{\rm NS}=1.85 \msun$ and a very massive companion star was found
$M_{\rm opt}=43 \msun$ (Kaper et al. 2006).  Since this is a
non-eclipsing system, a constraint may be put on the system
inclination resulting in the minimum mass of the optical companion at
$M_{\rm opt}=39 \msun$. We will use this lower limit as it provides
the best chance for the system survival (see below).  This system is a
potential progenitor of BH-NS binary. Since the NS is already in
place, the only requirement is that the companion forms a black hole
(BH) and the companion mass appears sufficient for BH
formation. Although, not well established, the transition from the NS
to BH formation may be argued to be anywhere in the range of the zero
age main sequence star mass $M_{\rm zams} = 20-40 \msun$ for solar
metallicity appropriate for this Galactic system.  Following the
recent calculations of compact object masses with the updated wind
mass loss we employ the transition at $M_{\rm zams} = 20 \msun$
(Belczynski et al. 2010a; Fryer et al. 2011). Therefore, if the
companion was a single star it would form a BH. However, this is a
secondary component in a binary that has not yet finished its
evolution. The primary, that obviously was initially a more massive
star in the binary, has already finished its evolution and formed a
NS. The claim that the secondary will form a BH seems
counter-intuitive. Yet, typical binary evolution leading to the
formation of X-ray binaries can easily explain such a system.

If initially this system hosted $40+30 \msun$ stars on a close orbit,
the $40 \msun$ star initiated Roche lobe overflow (RLOF) while still
on Main Sequence.  Since the mass ratio of both stars was close to
unity, the RLOF was dynamically stable and did not lead to the
development of a common envelope.  If this star lost more than $20
\msun$ and if de-rejuvenation was effective, the star transitioned
from the BH to NS formation zone, and became the currently observed
NS. The formation of a NS has set the current orbital parameters
including the high eccentricity as a result of natal kick the NS
received in an asymmetric supernova explosion (Hobbs et al. 2005).  If
part of the transferred material, say $10 \msun$ was successfully
accreted onto the companion it increased its mass to $40 \msun$,
becoming the currently observed optical star in GX 301-2. It is not our
intention to provide the detailed evolutionary history of this system
as we are much more interested in its future. However, we outlined the
likely formation scenario of this binary as it may seem curious to
form first a NS and then a BH. More detailed calculations and
discussion of similar processes are given by Wellstein \& Langer
(1999) and Belczynski \& Taam (2008b).

\begin{figure}
\includegraphics[width=1.0\columnwidth]{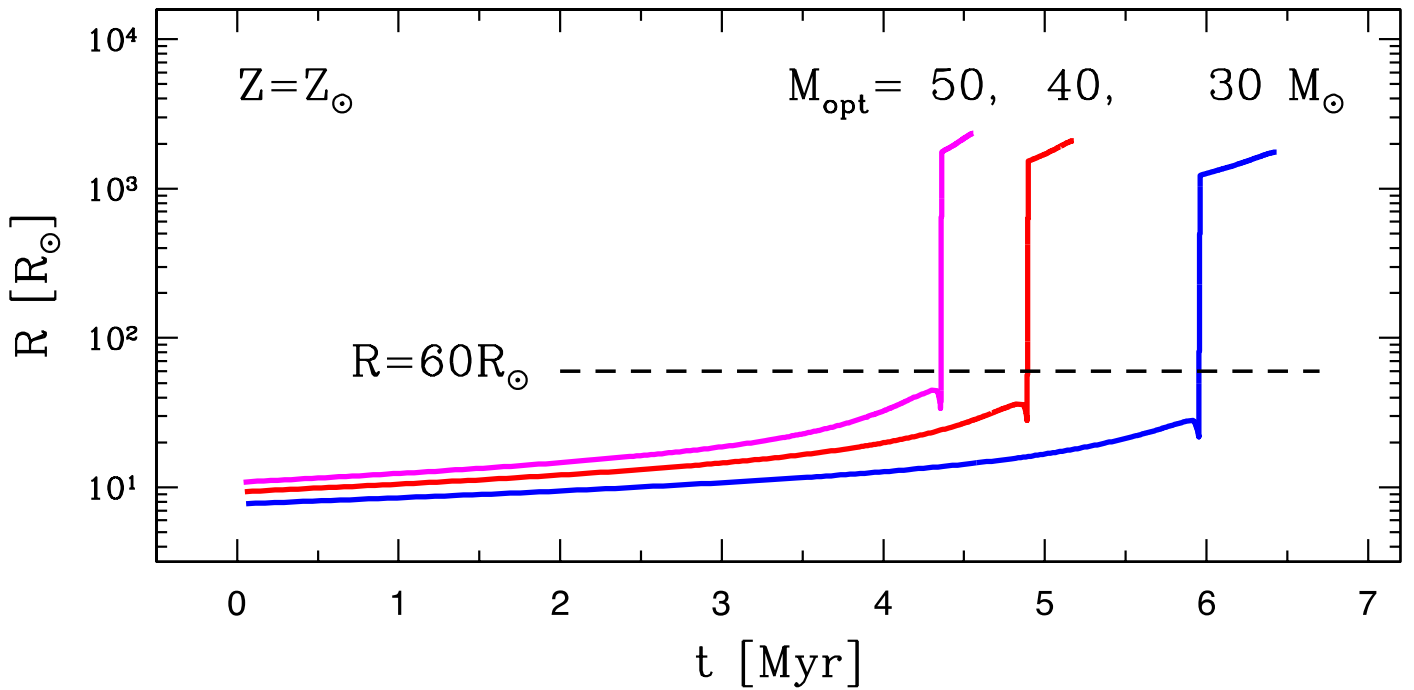}
\caption{Radius evolution of star with mass $30, 40, 50 \msun$ at
  solar metallicity obtained with Hurley et al. (2000) stellar
  evolutionary formulae.  At first, the star increases its radius by a
  factor of few over the long main sequence evolution. Then the
  expansion is extremely rapid and star increases its radius by factor
  of about hundred while crossing the Hertzsprung gap. Finally, after
  the ignition of helium in the core, the radial expansion slows down and
  eventually stops as star loses its entire H-rich envelope and
  becomes a compact naked helium star (not shown).  The optical component in
  GX 301-2 has $\sim 40 \msun$ and at periastron its Roche lobe radius
  is $\sim 60 \rsun$. The RLOF will start while the optical component
  is starting to cross Hertzsprung gap. This conclusion does not
  depend strongly on the optical component mass.}
\label{gx301}
\end{figure}

At the current orbital period the semi-major axis is $a=175\msun$,
while the closest encounter at periastron is $a_{\rm p}=95\rsun$. The
Roche lobe radius of the optical companion at periastron is $R_{\rm lobe}
= 60 \rsun$. The radius evolution of $30$, $40$ and $50 \msun$ stars
is shown in Figure~\ref{gx301}. It is apparent that the optical star
will eventually overfill its Roche lobe. From Figure~\ref{gx301} it
becomes clear that the RLOF will start shortly after the star leaves its main
sequence and begins crossing the Hertzsprung gap (HG). At the onset of
RLOF, the system is still highly eccentric as tidal forces are very
weak for such massive stars (e.g., Claret 2007). Although the
treatment of the mass transfer on an eccentric orbit is complicated
(e.g., Sepinsky et al. 2010), we can neglect this stage in the case of
GX 301-2. The expansion of the massive companion is so rapid that
within about $2000$ yr the star expands from $60$ to $200 \rsun$
engulfing entire system with its envelope (e.g., Hurley, Pols \& Tout
2000), initiating a common envelope phase (e.g., Webbink 1984). Once
the NS is within the stellar atmosphere of the companion the orbit
should rapidly circularize due to the increasing dynamical
friction. Since the entry into the atmosphere will occur somewhere
around periastron, it is natural to assume that the orbit will
circularize to a new separation equal to the size of pre-CE periastron
distance or maybe even smaller $a_{\rm pre,CE} \lesssim 95 \rsun$. As
an upper upper limit (albeit unphysical), we also consider the much
larger pre-CE orbit $a_{\rm pre,CE} \lesssim 175 \rsun$.  To calculate
the post CE separation, we use energy balance between the orbital
energy and the envelope binding energy. The binding energy is given by
$E_{\rm bind}=-(GM_{\rm opt}M_{\rm env})/(\lambda R_{\rm opt})$, where
$M_{\rm env}=25 \msun$ for a $M_{\rm opt}=40 \msun$ Hertzsprung gap
star and $\lambda$ is the parameter characterizing the mass
distribution within the star. The lower the $\lambda$ the harder it is
for a companion to eject the envelope. We have adopted $\lambda=0.2$
from Xu \& Li (2010) that is appropriate for massive Hertzsprung gap
star\footnote{We have obtained additional models from Xu \& Li as
  massive stars ($M>20\msun$) were not included in their original
  study.} of the relevant radius.

The final, post-CE, separation is determined by how much orbital
energy is required to eject the envelope.  If we assume all of the
orbital energy goes into unbinding the envelope, the post-CE
separations are found to be $a_{\rm post,CE}=0.28, 0.17 \rsun$ for
initial separations of $a_{\rm pre,CE}=175, 95 \rsun$, respectively.
Since the exposed core of the massive post main sequence star is a
massive helium star or Wolfe-Rayet star ($M_{\rm WR}=15\msun$) its
radius is $R_{\rm WR} \gtrsim 1\rsun$ (e.g., Petrovic, Pols \& Langer
2006). That means that the CE phase ended up with the NS sinking into
the helium core of the companion star, i.e. a merger. In a number of
population synthesis calculations, the parameter $\lambda$ denoting the
binding energy of the envelope is fixed (non-physically) to one
particular value for all stars of all masses and radii. Typically,
$\lambda=1.0$ was adopted due to lack of availability of more physical
values. If we use $\lambda=1.0$ instead of using calculated binding
energies, we effectively decrease the binding energy.  Nonetheless,
this system with $\lambda=1.0$ would still end up in a merger.  
Massive stars like the optical companion of GX 301-2 are not expected to
have $\lambda \gtrsim 1.0$ at any point during Hertzsprung gap (Xu \&
Li 2010), although it was recently claimed that the actual lambda
value may be $\sim 3-5$ times higher than currently estimated (Ivanova \&
Chaichenets 2011).  

For all the above estimates, we have assumed that all orbital energy
goes into unbinding the envelope (CE efficiency of $100\%$) thus
maximizing the survival chances of this system. Additionally, we have
neglected the internal structure of the companion star. Belczynski et
al. (2007) argued that since the Hertzsprung gap stars have not yet
developed clear core-envelope structure the CE phase always leads to a
merger if such a star is a donor independent of amount of orbital
energy available for the envelope ejection.  Even if some stars may
develop core-envelope structure later in the Hertzsprung gap and
survive CE, this is not likely in the case of GX 301-2 donor that will
initiate CE right after entering Hertzsprung gap (see
Fig.~\ref{gx301}).

Despite a number of very optimistic assumptions we have made to allow
for the formation of BH-NS binary, GX 301-2 will most likely end its
evolution as a single peculiar object formed in CE merger (a so-called
helium merger).  Several systems that we will discuss in the
subsequent sections will face a very similar fate, and we will provide
only basic information without the full description as it was done
here for GX 301-2.

\subsection{Vela X-1/GP Vel: CE merger}

\begin{figure}
\includegraphics[width=1.0\columnwidth]{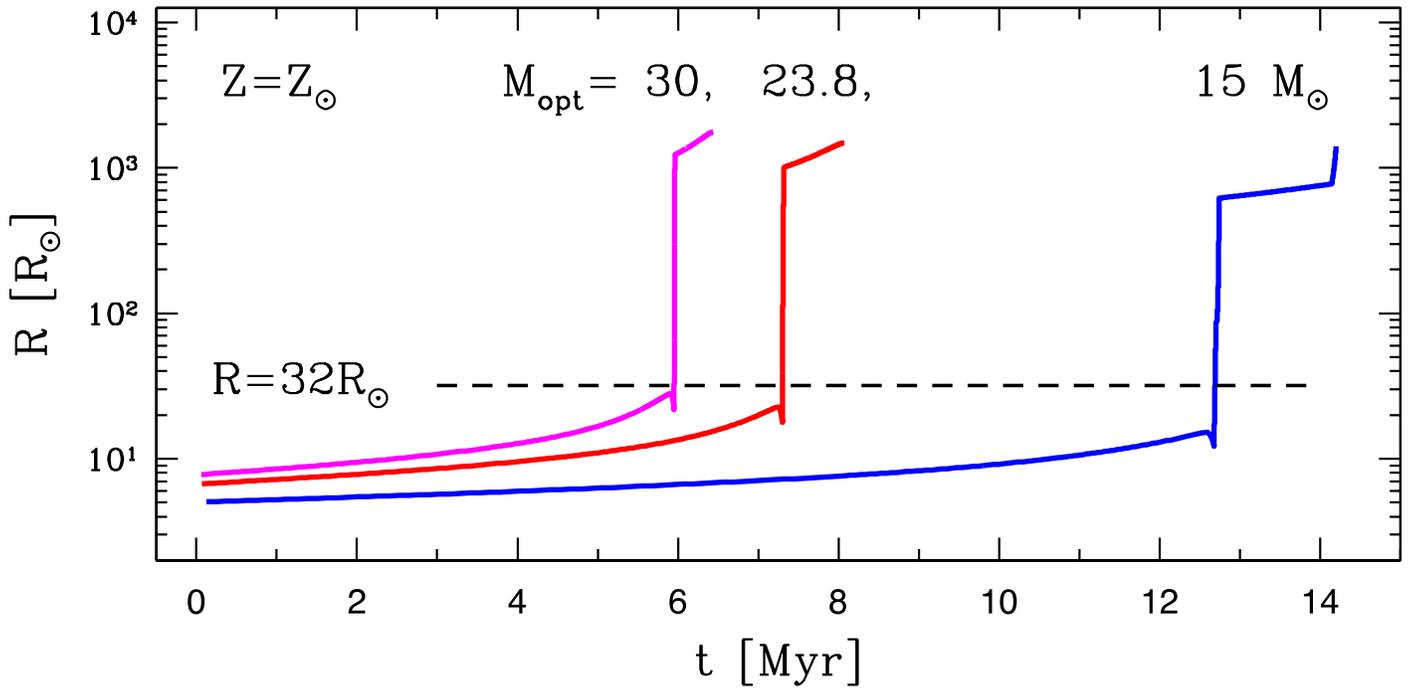}
\caption{Radius evolution of stars with mass $15, 23.8, 30 \msun$.
The optical component in Vela X-1 has mass $23.8 \msun$ and its Roche
lobe radius is $\sim 32 \rsun$. The RLOF will start while the optical
component is starting to cross Hertzsprung gap.} 
\label{vela}
\end{figure}

This system consists of a heavy NS ($M_{\rm NS}= 2.0 \msun$) orbiting
($P_{\rm orb}= 8.96$d) a massive companion ($M_{\rm opt} = 23.8
\msun$; e.g. Barziv et al. 2001) on a slightly eccentric orbit
($e=0.09$; e.g., Tomsick \& Muterspaugh 2010).  It is clearly seen
from Figure~\ref{vela} that the RLOF will start right after the
optical component enters Hertzsprung gap and it expands to reach its
Roche lobe radius of $R_{\rm lobe} =32 \rsun$. Due to the extreme mass
ratio of donor to accretor ($23.8:2$), the RLOF initiates a CE phase.
The current orbital separation is not subject to any strong changes
(small changes due to wind mass loss and tides) and we assume that the
orbital separation at the onset of CE is the same as it is today
$a_{\rm pre,CE} = 53 \rsun$. A $23.8 \msun$ star at the onset of
Hertzsprung gap has a core of about $7 \msun$ and massive $17 \msun$
envelope.  We optimistically assume that core-envelope structure is
well developed and system can survive CE phase provided that there is
enough orbital energy to expel the envelope.  The energy balance is
calculated for two values of $\lambda=0.65, 1.0$:  the first is a
physical value corresponding to the stellar structure of the donor and
the second is an upper limit on lambda for such a donor on the Hertzsprung
gap (Xu \& Li 2010).  The corresponding post-CE separations are
$a_{\rm post,CE} = 0.4, 0.6 \rsun$.  Since the core of the donor is a
massive helium star $M_{\rm WR} = 7\rsun$, its radius ($R_{\rm WR}
\gtrsim 1\rsun$; Petrovic et al. 2006) is larger than the post-CE
separation indicating that this system has not survived the CE phase,
forming another helium-merger.

\subsection{XTE J1855-026: CE merger}

This system is a $P_{\rm orb}=6.07$d eclipsing binary with a typical
NS ($M_{\rm NS}= 1.4 \msun$) and a massive companion ($M_{\rm opt} =
25 \msun$) with a negligibly small eccentricity ($e\lesssim 0.04$;
Corbet \& Mukai 2002, Tomsick \& Muterspaugh 2010). The corresponding
orbital separation is $a=42\rsun$ and the Roche lobe of the optical
companion $R_{\rm lobe} =26 \rsun$. The radius of the optical component
will reach its Roche lobe radius either at the very end of the main sequence
or at the very onset of Hertzsprung gap (similarity of Vela X-1
optical component allows to use Fig.~\ref{vela} to demonstrate this
point). The RLOF will lead to the development of the CE phase (note
the extreme mass ratio of donor to accretor $25:1.4$). If the star is
on main sequence during the onset of the CE phase, the binary
components will most certainly merge because, at this evolutionary stage,
there is no clear core-envelope structure. We assume that the optical
star in XTE J1855-026 is on the Hertzsprung gap while initiating CE
and that the system at least have a chance of survival. However, the
solution of energy balance leads to very small post-CE separations:
$a_{\rm post,CE} = 0.2, 0.3 \rsun$ for the realistic and limiting
values of $\lambda=0.7, 1.0$ respectively.  The massive helium core
($M_{\rm WR} = 7\rsun$) have radius ($R_{\rm WR} \gtrsim 1\rsun$;
Petrovic et al. 2006) exceeding the size of post-CE orbit.  This
system will merge in CE event, barring further binary evolution and 
potential double compact object formation.

\subsection{4U 1907+09: CE merger}

This system hosts a typical NS ($M_{\rm NS}= 1.4 \msun$) and a
massive companion ($M_{\rm opt} = 27 \msun$) on a rather close  
$P_{\rm orb}=8.38$d and eccentric $e=0.28$ orbit (Cox et al. 2005). 
Cox et al. (2005) estimated the companion radius at $R_{\rm opt} = 26
\rsun$ and provided arguments that the system is in wind-fed configuration
typical of HMXBs: {\em (i)} increase of X-ray luminosity at the periastron
passage and {\em (ii)} the radius of optical companion is well within its
Roche lobe even at periastron $R_{\rm lobe} = 31\rsun$.
However, they made an error in the Roche lobe radius estimate as the radius at 
the periastron for this particular binary configuration is actually much 
smaller $R_{\rm lobe} = 24.4\rsun$ (e.g., Eggleton 1983; Cox et al. (2005) 
made a typo in converting the Eggleton formula and their factor $q^{3/2}$ 
should actually read $q^{2/3}$). Therefore, if the radius estimate is
correct, the donor overflows its Roche lobe at periastron passages and the
increased periastron X-ray luminosity is due to a cyclic RLOF in addition 
to the wind accretion. If the companion radius is somewhat overestimated by
Cox et al. (2005) then companion may not be filling its Roche lobe, but it
is very close to RLOF at periastron. 

At solar metallicity, the radius of such a massive star reaches $\sim 27 \rsun$ 
at the end of the main sequence (MS), and expands to $\sim 1100 \rsun$ in the HG
(e.g., Hurley et al. 2000). Whether currently the system is going
through a cyclic and moderate RLOF at periastron passage or not, the
companion star is still in its MS because it has not yet expanded to
large post-MS evolution radii.  The expected near future expansion of
the companion on the HG will engulf the system in common envelope. Note
the extreme mass ratio (27:1.4). If the CE phase starts while the star
is still in its MS phase, the fate of the system is set and the binary
will merge. If the star initiates a CE phase while already on the HG,
the system is still most likely to end up in a merger (Belczynski et
al. 2007). However, we will ignore this fact for the sake of argument,
and we will show that simple energy balance estimation does not allow
this system survival in any case.

The CE starts once the optical companion begins its rapid expansion at the 
onset of the HG phase. Whether we assume the periastron distance as the 
initial CE separation or actual (most likely overestimated) binary separation 
$a_{\rm pre,CE} = 39, 54 \rsun$, the system merges during CE. The post-CE
separation obtained from the energy balance is calculated to be 
$a_{\rm post,CE} = 0.20, 0.26 \rsun$, resulting in $R_{\rm lobe} = 0.10, 0.14 
\rsun$ while the massive WR star ($8 \msun$ He core of the optical companion) 
radius is $R_{\rm WR} = 0.8\rsun$ for both initial separation cases, respectively. 
Again we have used realistic $\lambda=0.7$ appropriate for such massive star
at its given radius, and we have assumed conservatively the 100\% efficiency of 
orbital energy conversion into unbinding the envelope. 

In all cases, this system will end in a CE merger.

\subsection{Cir X-1: CE merger}

This binary was at first believed to host a BH due to the presence of jets.
However, the discovery and then later confirmation of thermonuclear bursts
has shown beyond a doubt that the compact object is a NS (Lineares et al. 2010). 
The orbital period ($P_{\rm orb}=16.6$d) and eccentricity are well established 
($e=0.45$). However, due to the high extinction toward Cir X-1 and potential
large contribution of a disk emission in UV, optical and IR the mass and
evolutionary status of a companion are not well constrained. Estimates range
from $3-5 \msun$ subgiant (Johnston, Fender \& Wu 1999) to $10 \msun$
suprgiant (Jonker, Nelemans \& Bassa 2007). 

If Cir X-1 were hosting a BH as it was originally suspected and if the
companion was massive enough to form a NS (e.g., $\sim 10 \msun$) then it
would be a potential progenitor of BH-NS system. However, this option is no 
longer supported. This peculiar XRB is a potential progenitor of either a 
NS-WD system (low mass $3-5 \msun$ companion) or NS-NS system (high mass
companion $10 \msun$). In the following estimates we assume a canonical NS
mass of $1.4 \msun$ for the compact star in Cir X-1, and adopt either a low
or high mass companion $M_{\rm opt} = 4, 10 \msun$. 

For the low mass companion, the RLOF commences when the star is on HG and
leads most likely to CE evolution (mass ratio of 4:1.4) that ends up in a 
merger with a NS. The orbital separation just prior RLOF is $a=48 \rsun$ with 
companion radius $R_{\rm opt}=12 \rsun$. We adopt the periastron distance 
for the onset of CE  $a_{\rm pre,CE}=26 \rsun$ with $e=0.45$ as the tidal
forces were not able to circularize the system for this (radiative) companion 
(e.g., Claret 2007). 
We calculate CE with $\lambda=0.5$ (Xu \& Li 2010) and obtain a final orbital
separation $a_{\rm post,CE}=0.21 \rsun$. The companion lost its entire
envelope and will become a low mass helium star $M_{\rm WR}=0.6 \msun$ with
radius $R_{\rm WR}=0.14 \rsun$ and thus it exceeds its new Roche lobe 
$R_{\rm lobe} = 0.07 \rsun$. The NS has inspiraled into He core of the 
$4 \msun$ HG companion. 

For the high mass companion, the RLOF will start much earlier, but again when
the donor star is crossing HG. The initial periastron separation is $a_{\rm
pre,CE}=63 \rsun$ with $e=0.45$ and the donor radius $R_{\rm opt}=19 \rsun$. 
We use $\lambda=0.7$ appropriate for this given donor and we calculate the 
final orbital separation $a_{\rm post,CE}=0.26 \rsun$. The companion 
became a helium star with mass $M_{\rm WR}=2.1 \msun$ and with
radius $R_{\rm WR}=0.35 \rsun$ and thus it exceeds its new Roche lobe  
$R_{\rm lobe} = 0.11 \rsun$. The NS has inspiraled into He core of the 
$10 \msun$ HG companion.

\subsection{LMC X-1:  CE merger}

This is the first XRB discovered in the Large Magellanic Cloud
(LMC). The system parameters are relatively well established (Orosz et
al. 2009; Ziolkowski 2011).  The compact object is a BH with mass
$M_{\rm BH}=10.9 \msun$ in an orbit around a main sequence star with
mass $M_{\rm opt}=31.8 \msun$ and radius $R_{\rm opt}=17.0
\rsun$. The orbit is circular or if there is any eccentricity it is
negligibly small ($e \lesssim 0.03$). The orbital period is found to
be $P_{\rm orb}=3.91$d corresponding to an orbital separation $a=36.5
\rsun$.  This indicates that the optical star almost fills its Roche
lobe ($R_{\rm lobe}=17.3 \rsun$).  Since the mass of a optical star is
above the BH formation limit for single stellar models and for our
adopted evolutionary scenario (Belczynski et al.  2002; 2008a), this
system is a candidate progenitor of BH-BH system.  If binary
interactions can remove a significant part of the MS star mass, it could
also potentially form BH-NS system.

The initial mass of the optical companion was estimated at about $M_{\rm opt,zams}=35 \msun$ 
(Orosz et al. 2009; Ziolkowski 2011).  The radius evolution of a $M_{\rm zams}=35
\msun$ star at $Z=0.3\zsun$ metallicity typical of the LMC is shown in Figure 3. 
For $M_{\rm zams}=35 \msun$ we see that the radius of the star reaches its Roche
lobe at about $5$ Myr of MS evolution. For this given set of orbital parameters,
evolutionary stage of a donor and high mass ratio of the binary, 
(31.8:10.9) the RLOF will develop into a CE phase. The CE evolution is
bound to lead to the merger (MS donor) and the BH will sink into the He
enriched core of MS star. At the time of merger, the donor should have
evolved through about $90\%$ of its MS lifetime. The core mass at the end of
MS is similar to that at the beginning of HG: $\sim 11 \msun$ for the
optical star in LMC X-1.

\subsection{LMC X-3:  BH-WD Binary}

This system is another black hole binary in the LMC. The optical companion was established 
to be a BV star with mass $4-8 \msun$ and radial velocity curve with $P_{\rm orb} 
= 1.7$d and potential small eccentricity $e=0-0.1$ was obtained by Cowley et al. 
(1983). The lack of eclipses ($i<70\deg$) places a lower limit only  
on the mass of compact object, but it is high enough to indicate a black hole
binary. If the maximum mass of the companion ($8\msun$) is confirmed, this
system is a potential BH-NS system progenitor. However, recent estimates
indicate that LMC X-3 will face a different fate. 

Soria et al. (2001) study included effects of irradiation, and they estimated 
that the companion is a subgiant star filling its Roche lobe B5 IV 
($M_{\rm opt}=4.7\msun$) and obtained limit for the mass of a black hole 
$M_{\rm BH} > 5.8 \msun$. 
This system is already at RLOF and a companion star is losing mass ($R_{\rm lobe}=4.7\rsun$ 
if we adopt a BH mass $M_{\rm BH} =5.8 \msun$).
The evolutionary state of the companion to put it at a right size to commence 
RLOF at the observed orbital period is either late MS or early HG. The
helium core of such a donor will be found at $0.5-0.7 \msun$. The RLOF will
continue through HG, Red Giant Branch to stop finally at Core Helium Burning
at which stage the companion will contract. The remaining envelope of
companion will be lost in stellar wind exposing $0.6-0.8 \msun$ CO white
dwarf. The orbit, due to mass transfer (from less massive companion to more 
massive BH) will expand to over 100 days forming a wide (non-coalescing) BH-WD
binary.

\subsection{M33 X-7:  CE Merger} 

This system is a massive binary consisting of a black hole $M_{\rm
  BH}=15.65\msun$ and an O7-8 III companion $M_{\rm opt}=70\msun$ on a
close and almost circular orbit: $P_{\rm orb}=3.45$d, $e=0.0185$
(Orosz et al. 2007).  Following Valsecchi et al. (2010) we adopt M33
metallicity of $Z=0.01$ or $50\%$ solar.

The radius of the optical component is estimated to be $19.6\rsun$
(Orosz et al.  2007).  The orbital separation is found to be 
$a=42.4\rsun$ and the Roche lobe radius at periastron for the optical
component is $R_{\rm lobe}=21.3\rsun$. The companion is still in
its MS and almost fills its Roche lobe. It will take only about $\sim 0.2$
Myr for such a massive star to expand and overfill its Roche lobe. Due
to the extreme mass ratio ($70:15.65$), the system will enter the
common envelope phase that will lead to a merger (MS donor). The
massive BH will sink into the core of the companion: at this point the
companion has evolved through about $60-70\%$ of its MS lifetime and
the core will mostly consist of helium nuclei.

\section{Discussion}

Including our previous studies we have shown that out of $11$ very massive 
HMXB with reasonably known binary parameters: $2$ (IC10 X-1, NGC300 X-1) will 
form close BH-BH systems, $1$ (Cyg X-1) will have a $30\%$ chance to form a
BH-NS system (with $1\%$ probability of this being a close system), $1$ (LMC
X-3) will form a wide BH-WD system, and the remaining $7$ binaries will
merge in common envelope (GX 301-2, Vela X-1, XTE J1855-026, 4U 1907+09, Cir
X-1, LMC X-1, M33 X-1). The summary of our results is given in Table~1.

In the helium merger model of long-duration gamma-ray bursts (GRBs) the compact
object, either a NS or a BH, sinks into the helium rich core of its
companion. A compact object accretes at a high rate $\gtrsim 0.01
\mps$ (Fryer \& Woosley 1998, Zhang \& Fryer 2001) and, in the NS case, a BH forms. The
helium core as it accretes onto a BH forms a transient torus (Barkov
\& Komissarov 2011) leading to a standard scenario for a GRB engine
(Popham, Woosley \& Fryer 1999).  Helium merger configuration is most
easily provided in the common envelope merger of HMXBs in which an
optical companion is already an evolved star beyond main sequence (He
core) or at the end of main sequence (high concentration of He in the
core). In the merger, part of a companion star is ejected (but not
fully) away from the system, and the compact object sinks into
helium-rich core.  The so-called ``Christmas burst'' (GRB 121225) may
be potentially the first identifiable example of a helium merger GRB,
with indication of an inner compact source and a surrounding shell
(Thone et al. 2011).

Fryer, Belczynski \& Thone (2013) have employed theoretical population
synthesis predictions for common envelope mergers and estimated the
characteristic observational features of helium merger GRBs. Their
model calculates a GRB luminosity based on a helium core mass
(required to be typically larger than $4 \msun$) and establishes the
position of the shell surrounding the central GRB engine based on the radius
of the companion star (CE donor). Here we use our empirically based
estimates and employ Fryer et al.  (2013) model to estimate typical
luminosities of helium merger GRBs.

In the case of GX 301-2, we expect a NS to sink into a $60 \rsun$ Hertzsprung gap 
star with $15 \msun$ helium core and GRB luminosity of $L_{\rm grb} =
10^{49}, 10^{51}$ erg s$^{-1}$ for neutrino annihilation and Blandford-Znajek
emission model respectively (for each system, we will provide two 
values for $L_{\rm grb}$ corresponding to these two emission mechanisms).
For Vela X-1 a NS will enter a $32 \rsun$ Hertzsprung gap star with $7 \msun$ 
helium core and  $L_{\rm grb} = 10^{46}, 10^{50}$ erg s$^{-1}$.
In binary XTE J1855-026 a NS will merge into $26 \rsun$ Hertzsprung gap star
with $7 \msun$ helium core and $L_{\rm grb} = 10^{46}, 10^{50}$ erg s$^{-1}$. 
In case of 4U 1907+09 we predict a merger of a NS and a $26 \rsun$ late main 
sequence or early Hertzsprung gap star with $8 \msun$ helium-rich core
and $L_{\rm grb} = 10^{47}, 5 \times 10^{50}$ erg s$^{-1}$. 
For Cir X-1 a NS will sink into a $12-19 \rsun$ Hertzsprung gap star 
with $0.6-2.1 \msun$ helium core.  For this case, a GRB is not expected but 
such mergers may account for some subset of ultraluminous supernovae (Fryer et al. 2013). 
For LMC X-1 a BH will merge into a $\sim 11 \msun$ helium-rich core of a 
$17.3 \rsun$ star at the end of its main sequence evolution with a potential
GRB at $L_{\rm grb} = 10^{48}, 10^{51}$ erg s$^{-1}$. 
A BH in M33 X-1 will sink into a $21.3 \rsun$ main sequence companion that 
has evolved through about $60-70\%$ of its central hydrogen burning. Since 
the optical companion has mass of $\sim 70\msun$ its helium-rich core is also 
very massive $\sim 20 \msun$ and potential GRB luminosity is very high 
$L_{\rm grb} = 10^{51}, 5 \times 10^{51}$ erg s$^{-1}$.

There are at least 4 HMXB binaries in our Galaxy with very massive
donors ($M_{\rm opt}>20 \msun$; required for a GRB) that will merge in
the soon-to-come common envelope phase (see Tab.~1). As discussed
above the merger will be followed shortly by a gamma ray burst (Fryer
\& Woosley 1998; Thone et al. 2011). The HMXB lifetime may be
estimated from the lifetime of an optical component. For the
considered binaries (GX 301-2, Vela X-1, XTE J1855-026, 4U 1907+09)
the HMXB lifetime can be roughly estimated as the difference between
the lifetime of the companion star and the minimum time required to
form a NS i.e. $5$ Myr (Belczynski \& Taam 2008).  The companion
lifetime depends on its mass and it is $5.3$ Myr for a $43\msun$ star
(GX 301-2), $8$ Myr for $24\msun$ star (Vela X-1), $7.8$ Myr for
$25\msun$ star (XTE J1855-026), and $7.3$ Myr for $27 \msun$ star (4U
1907+09). This leads to the X-ray lifetime estimates of $\tau=$ $0.3$,
$3$, $2.8$ and $2$ Myr, respectively. The expected formation rate of
such X-ray binaries is $R = \sum \tau_i^{-1}=4.5$ Myr$^{-1}$. Since 
the HMXB lifetimes are very short, this is also the estimate of the 
helium merger Galactic GRB rate. The rate estimate is formally 
dominated by the contribution of GX 301-2 and in this case this system
deserves a few more comments. The companion in GX 301-2 might have 
gained up to $20 \msun$ from the NS progenitor, and 
the rejuvenation might not have been complete. Thus it is quite likely
that the expected lifetime of GX 301-2 as an HMXB is longer, up to $3$
Myr. If we adopt such long lifetime the formation rate of helium
merger GRBs drops to $1.5$ Myr$^{-1}$.  These rates have been
calculated assuming that the entire Galaxy has been searched for HMXB
and our observational sample is complete. In reality HMXBs have been
found only in about $25\%$ of the volume of the disk (e.g., Ozel et
al. 2010). Thus the estimated rates have to be corrected for that and
are in the range $R_{\rm Galactic}=6-18$ Myr$^{-1}$.

The star formation in the Galactic disk is approximately 
constant and at the level of $3.5 \mpy$ (e.g., O'Shaughnessy et al. 2006). 
The rate of helium merger GRBs may
be translated into ${\cal R}_{grb} \approx 1.6 - 5.1 \times 10^{-6} \msun^{-1}$ and
is expressed per unit of star forming mass. This value can be used to estimate 
the GRB rate in other galaxies or in cosmological population studies. 
Since the delay of helium CE
merger events is rather short ($\lesssim 10$ Myr), these particular GRBs are 
expected in any galaxy with ongoing star formation. 

It must be noted that our estimate provides only a lower limit on the helium
merger GRB rate. There are a number of HMXBs in our Galaxy without fully
established orbital/component parameters (Liu, van Paradijs \& van den
Heuvel 2006), some of which may be potential GRB progenitors. 
There may be ways to produce a helium merger GRB without a HMXB phase, for
example NS ejection into a companion via supernova natal kick (e.g., Fryer 
et al. 2013). 

Nevertheless, our estimate provides a rather significant number of GRBs in the
history of the Galactic disk $\sim 60,000-180,000$ (for $10$ Gyr of constant star
formation) and we check whether such a GRB may have affected life on Earth.   
The extent of the disk, where most of star formation is taking
place, may be approximated by radius of $18$ kpc and vertical extent of
$0.60$ kpc (Paczynski 1990) and that gives disk volume of $V_{\rm disk}
\approx 610$ kpc$^3$. Long GRBs with their maximum energy output of 
$5 \times 10^{51}$ erg/s lasting $\sim 10$s may have a big impact or are 
claimed to be lethal for Earth biosphere if exploding within $2$ kpc 
(Nakar 2010). The volume of impact sphere within the disk 
 is then about $V_{\rm impact} \approx 7.5$ kpc$^3$. 
And therefore the Galactic rate needs to be reduced by factor of
$\sim 80$ ($V_{\rm disk}/V_{\rm impact}$) to provide the rate estimate 
of GRBs in the impact zone: $0.073-0.22$ Myr$^{-1}$. The long GRB power is collimated 
in a relativistic jet with beaming of $\sim 5^{\circ}$ (e.g., Frail et al. 2001; 
Soderberg et al. 2006). 
Since only a direct hit on the Earth biosphere may be lethal or cause a major 
mass extinction, this leads to further reduction by factor of $\sim 500$. 
Finally, we obtain the rate of direct GRB hits coming from the impact zone 
at the level $0.15-0.45$ Gyr$^{-1}$. Over a $4.5$ Gyr of Earth history the expected
number of direct nearby hits is in the range  $0.65 - 2.0$. 
There are at least five major mass extinction events noted in the last $0.5$ Gyr 
in the marine data (Alroy 2008). The probability of direct impact from  a helium 
merger GRBs in this period is in the range $\sim 7.2 - 22\%$). Thus it is quite 
possible that one of these events has been caused by a nearby helium GRB.

\acknowledgements Authors acknowledge support from MSHE grants N203
404939 (KB) and NASA Grant NNX09AV06A to the UTB Center for
Gravitational Wave Astronomy (KB).  The work of C.F. was funded by the
Department of Energy, and supported by its contract W-7405-ENG-36 to
Los Alamos National Laboratory and by the National Science Foundation
under Grant No. NSF PHY11-25915.

\begin{deluxetable}{llllccl}
\tablewidth{450pt}
\tablecaption{Fate of HMXBs}
\tablehead{No & Name & $M_{\rm x}/\msun$ & $M_{\rm opt}/\msun$ & $P_{\rm orb}$ & $e$ & Fate}
\startdata
1  & IC10 X-1      & $30$ (BH)    & $25$ (WR)   & $5.6$h  & $0$     & close BH-BH ($100\%$) \\
2  & NGC300 X-1    & $20$ (BH)    & $15$ (WR)   & $6.7$h  & $0$     & close BH-BH ($100\%$) \\
3  & Cyg X-1       & $15$ (BH)    & $19$ (O)    & $5.6$d  & $0$     & close BH-NS ($\lesssim 1\%$) \\
   &               &              &             &         &         &           \\
4  & GX 301-2      & $1.9$ (NS)   & $43$ (O)    & $41.5$d & $0.46$  & CE merger \\
5  & Vela X-1      & $2.0$ (NS)   & $24$ (O)    & $8.96$d & $0.09$  & CE merger \\
6  & XTE J1855-026 & $1.4$ (NS)   & $25$ (O)    & $6.07$d & $0.04$  & CE merger \\
7  & 4U 1907+09    & $1.4$ (NS)   & $27$ (O)    & $8.38$d & $0.28$  & CE merger \\
8  & Cir X-1       & ??\tablenotemark{a}\ \ (NS) & $3-5$ (A/B) & $16.6$d & $0.45$ & CE merger \\
   &               &              & $10$ (O)\tablenotemark{b}       &         &        & CE merger \\
   &               &              &             &         &         &  \\
9  & LMC X-1       & $10.9$ (BH)  & $31.8$ (O)  & $3.91$d & $0.0$   & CE merger \\
10 & LMC X-3       & $>5.8$ (BH)  & $4.7$  (B)  & $1.7$d  & $<0.1$ & wide BH-WD \\
11 & M33 X-1       & $15.65$ (BH) & $70$   (O)  & $3.45$d & $0.019$ & CE merger \\

\enddata
\label{hmxb}
\tablenotetext{a}{
It is established that compact object in this system is a NS, however its
mass has not been measured.  In our calculations, we have assumed a NS mass of
$1.4 \msun$.}
\tablenotetext{b}{
The mass and evolutionary status of the companion is not well established. 
We use the two different mass estimates, however the fate of the system 
seems insensitive to these uncertainties. 
}

\end{deluxetable}
\clearpage

\clearpage

\clearpage

\begin{figure}
\includegraphics[width=1.0\columnwidth]{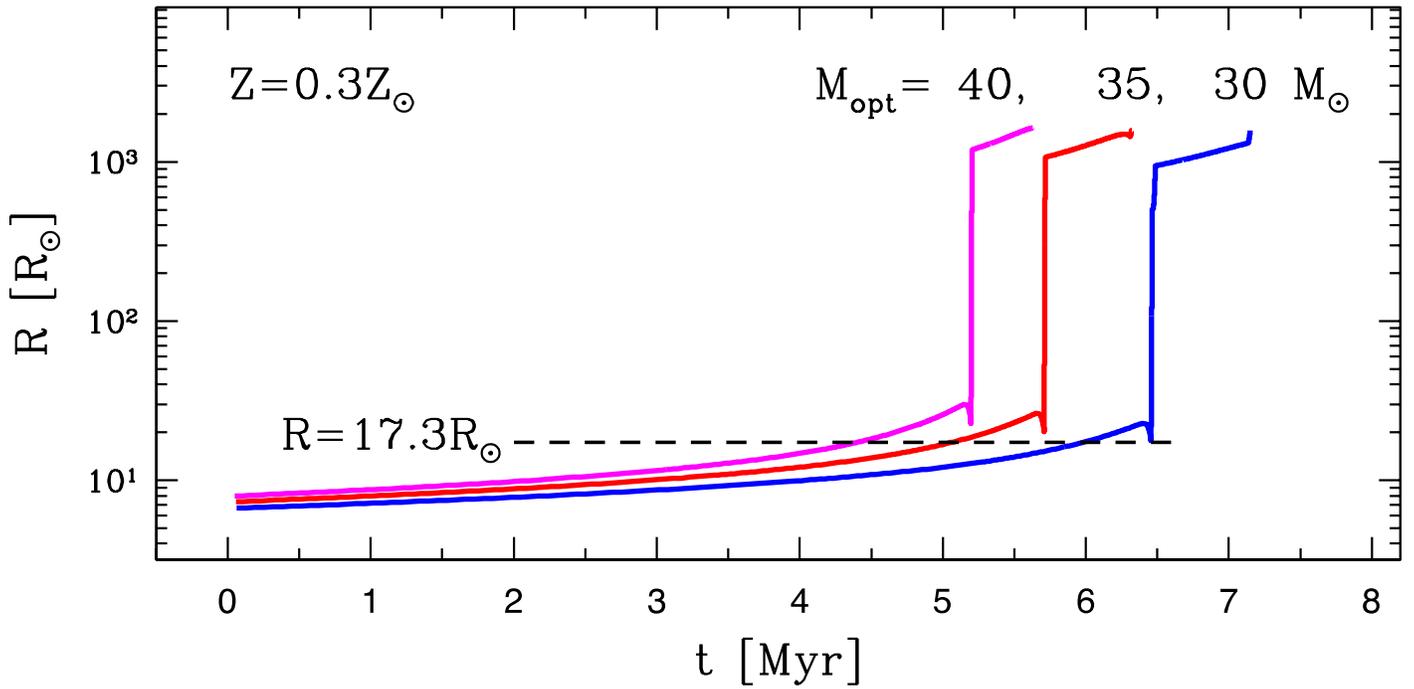}
\caption{Radius evolution of stars with mass $30, 35, 40 \msun$ for a small
metallicity typical of the LMC.
Initial mass (ZAMS) of the LMC X-1 optical component was estimated at $35
\msun$. The current mass and radius are $31.8 \msun$ and $17.0 \rsun$,
respectively. The estimated Roche
lobe radius is $\sim 17.3 \rsun$. The RLOF will start very soon ($\lesssim
10^5$ yr) while the optical component is still on Main Sequence.} 
\label{lmcx1}
\end{figure}

\end{document}